\documentclass[11pt]{article}
\usepackage{amsmath,amssymb}
\usepackage{graphicx}
\usepackage{amsfonts}

\usepackage{amsthm}

\usepackage[USenglish]{babel} 
\usepackage[T1]{fontenc}
\usepackage[ansinew]{inputenc}
\usepackage{mathrsfs}
\usepackage{stmaryrd} 
\usepackage{epstopdf} 
\usepackage{verbatim}
\usepackage{MnSymbol}  
\usepackage{accents}   
\usepackage{bbm}			 
\usepackage{color}     
\usepackage{authblk}   

\numberwithin{equation}{section}
\numberwithin{figure}{section}

\title{Symplectic model reduction methods for the Vlasov equation}
\date{}
\author[1,2]{Tomasz M. Tyranowski\thanks{\texttt{tomasz.tyranowski@ipp.mpg.de}}}
\author[1,2]{Michael Kraus\thanks{\texttt{michael.kraus@ipp.mpg.de}}}

\affil[1]{\small Max-Planck-Institut f\"ur Plasmaphysik \authorcr Boltzmannstra{\ss}e 2, 85748 Garching, Germany\vspace{.5em}}
\affil[2]{\small Technische Universit\"{a}t M\"{u}nchen, Zentrum Mathematik \authorcr Boltzmannstra{\ss}e 3, 85748 Garching, Germany\vspace{.1em}}

\begin{document}

\maketitle

\begin{abstract}
Particle-based simulations of the Vlasov equation typically require a large number of particles, which leads to a high-dimensional system of ordinary differential equations. Solving such systems is computationally very expensive, especially when simulations for many different values of input parameters are desired. In this work we compare several model reduction techniques and demonstrate their applicability to numerical simulations of the Vlasov equation. The necessity of symplectic model reduction algorithms is illustrated with a simple numerical experiment.
\end{abstract}

\section{Introduction}
\label{sec:intro}

Kinetic models like the Vlasov equation provide the most accurate description for a plasma beyond the interaction of individual particles.
While physically comprehensive, such models are often too expensive to be solved numerically under realistic conditions, especially in many-query contexts like uncertainty quantification, inverse problems or optimization.
In such situations reduced complexity models can provide a good compromise between computational cost and physical completeness of numerical models, and facilitate the solution of problems that are otherwise unfeasible.

A common approach for the construction of reduced complexity models are modal decomposition techniques such as proper orthogonal decomposition (POD, see \cite{Antoulas2001}), also known as principal component analysis (PCA, see \cite{Moore1981}), and dynamic mode decomposition (DMD, see \cite{AllaKutz2017}).
Here, existing complex models are replaced by reduced models, which preserve the essential features of the original systems, but require less computational effort.

Algorithms obtained by such model order reduction techniques usually consist of two stages: an offline stage, where the reduced basis is constructed from empirical or simulation data of a physical system, referred to as snapshots, and an online stage, where the system is solved in the reduced basis. These techniques use singular-value decomposition (SVD) to identify dominant global modes in the snapshots, in such a way that the basis constructed from these modes spans the data optimally. Standard Galerkin projection methods are used to obtain approximate operators on such a basis (see \cite{Kunisch2002}). This approach is often amended by techniques such as the discrete empirical interpolation method to allow for efficient evaluation of nonlinearities (see \cite{Chaturantabut2010}).

While model order reduction has been successfully applied to finite difference and finite volume discretizations of fluid equations (see \cite{Lassila2014} for an overview), its applicability is not yet well studied for kinetic equations.
Thus the main goal of this work is to demonstrate the usefulness of model reduction techniques for numerical simulations of the Vlasov equation using particle methods.
Further, we will show that preserving the Hamiltonian structure of the system in the reduction procedure is key to obtaining stable and accurate reduced models.

\subsection{Particle methods for the Vlasov equation}

In this work we consider the Vlasov equation

\begin{equation}
\label{eq: Vlasov equation}
\frac{\partial f}{\partial t} + v \frac{\partial f}{\partial x} - E(x) \frac{\partial f}{\partial v} = 0,
\end{equation}

\noindent
for the particle density function $f=f(t,x,v)$ of plasma consisting of charged particles of unit mass and  unit negative charge, where $E(x) = -\frac{\partial \phi}{\partial x}$ is an external electrostatic field with the potential $\phi=\phi(x)$, and $x$ and $v$ are vectors in $\mathbb{R}^d$ with $d=1,2,3$. The standard approach to particle  methods consists of the Ansatz

\begin{equation}
\label{eq: Ansatz for f}
f(t,x,v) = \sum_{i=1}^{n} w_i \delta(x-X_i(t)) \delta(v - V_i(t))
\end{equation}

\noindent
for the particle density function, where $X_i(t)$ and $V_i(t)$ represent the position and velocity of the $i$-th particle, respectively, and $w_i$ its weight. Substituting \eqref{eq: Ansatz for f} in \eqref{eq: Vlasov equation}, one obtains a system of ordinary differential equations (ODEs) for $X_i(t)$ and $V_i(t)$, namely

\begin{align}
\label{eq: ODEs for the particles}
\dot X_i &= V_i, \nonumber \\
\dot V_i &= \frac{\partial \phi}{\partial x}(X_i), \qquad\qquad i=1,\ldots,n.
\end{align}

\noindent
It can be easily verified that \eqref{eq: ODEs for the particles} has the form of a Hamiltonian system of equations 

\begin{align}
\label{eq: Hamiltonian ODEs for the particles}
\dot X_i &= \frac{\partial H}{\partial V_i}, \nonumber \\
\dot V_i &= -\frac{\partial H}{\partial X_i}, \qquad\qquad i=1,\ldots,n,
\end{align}

\noindent
with the Hamiltonian $H$ given by

\begin{equation}
\label{eq: Hamiltonian for the particles}
H(X,V) = \sum_{i=1}^n \Big[\frac{1}{2} V_i^2 - \phi(X_i) \Big],
\end{equation}

\noindent
where $X = (X_1, \ldots, X_n)$ and $V = (V_1,\ldots, V_n)$ are vectors in $\mathbb{R}^{nd}$.  Note that, for simplicity and brevity, we express the Hamiltonian in terms of the velocity vector $V$ rather than the canonical conjugate momentum vector $P$, since for a fixed\footnote{In the general case of self-consistent electromagnetic fields we have that $P_i = mV_i+qA(X_i,t)$, where $A$ is the vector potential of the electromagnetic field, and $m$ and $q$ are the mass and charge of the particles (see, e.g., \cite{JacksonElectrodynamics}, \cite{MarsdenRatiuSymmetry}).} external electric field and particles of unit mass we have $P_i=V_i$.

\subsection{Geometric integration}
The Hamiltonian system \eqref{eq: Hamiltonian ODEs for the particles} possesses several characteristic properties. Its flow $F_t: \mathbb{R}^{2nd}\longrightarrow \mathbb{R}^{2nd}$ preserves the Hamiltonian, i.e. the total energy, as well as the canonical symplectic form $\omega = \sum_{i=1}^n \sum_{j=1}^d dX_i^j \wedge dV_i^j$. The latter property expressed in terms of the standard basis for $\mathbb{R}^{2nd}$ takes the form of the condition

\begin{equation}
\label{eq: Symplecticity of the flow}
(DF_t)^T \mathbb{J}_{2nd} DF_t = \mathbb{J}_{2nd},
\end{equation} 

\noindent
where $DF_t$ denotes the Jacobi matrix of the flow map $F_t$, $\mathbb{J}_{2nd}$ denotes the canonical symplectic matrix defined as

\begin{equation}
\label{eq: Canonical symplectic matrix}
\mathbb{J}_{2nd}=\left(\begin{matrix}
0 & \mathbb{I}_{nd} \\
-\mathbb{I}_{nd} & 0
\end{matrix}\right),
\end{equation}

\noindent
and $\mathbb{I}_{nd}$ is the $nd \times nd$ identity matrix (see, e.g., \cite{HLWGeometric}, \cite{HolmGMS}, \cite{MarsdenRatiuSymmetry}).

In principle, general purpose numerical schemes for ODEs can be applied to Hamiltonian systems such as \eqref{eq: ODEs for the particles}. However, when simulating these systems numerically, it is advisable that the numerical scheme also preserves geometric features such as symplecticity \eqref{eq: Symplecticity of the flow}. Geometric integration of Hamiltonian systems has been thoroughly studied (see \cite{HLWGeometric}, \cite{HallLeokSpectral}, \cite{KaneMarsden2000}, \cite{KrausPHD}, \cite{LeokZhang}, \cite{MarsdenPatrickShkoller}, \cite{MarsdenWestVarInt}, \cite{McLachlanQuispel}, \cite{OberBlobaum2015}, \cite{RowleyMarsden}, \cite{SanzSerna}, \cite{TyranowskiPHD}, \cite{TyranowskiDesbrunRAMVI}, \cite{TyranowskiDesbrunLinearLagrangians} and the references therein) and symplectic integrators have been shown to demonstrate superior performance in long-time simulations of such systems, compared to non-symplectic methods. Long-time accuracy and near preservation of the Hamiltonian by symplectic integrators have been rigorously studied using the so-called backward error analysis (see, e.g., \cite{HLWGeometric} and the references therein).  Application of geometric integration to particle-in-cell (PIC) simulations of the Vlasov equation coupled to self-consistent electromagnetic fields satisfying the Maxwell equations (i.e., the Vlasov-Maxwell system) was proposed in \cite{KrausGEMPIC}, \cite{Qin2016}, \cite{XiaoLiuQin2013}, \cite{XiaoQinLiu2018}, \cite{Xiao2015}. 

\subsection{Symplectic model reduction}

For the aforementioned reasons it appears desirable to preserve the Hamiltonian structure also in model reduction. In fact, it has been found that preserving the Hamiltonian structure in the construction of the reduced spaces preserves stability \cite{AfkhamHesthaven2017}, which is not guaranteed using non-structure-preserving model reduction techniques (\cite{Prajna2003}, \cite{RathinamPetzold2003}).
To this end, standard model reduction techniques such as proper orthogonal decomposition have been modified towards the so-called proper symplectic decomposition (\cite{PengMohseniProceedings2016}, \cite{PengMohseniPreprint2016}, \cite{PengMohseni2016}), which does indeed preserve the canonical symplectic structure of many Hamiltonian systems in the reduction procedure. Similarly, greedy algorithms (\cite{AfkhamHesthaven2017}, \cite{AfkhamHesthaven2018}, \cite{AfkhamHesthaven2019}) can be used to construct the reduced basis in a Hamiltonian-structure preserving way, and recently also non-orthonormal bases have been considered \cite{Buchfink2019}, showing improved efficiency over orthonormal bases. See also \cite{Carlberg2015}, \cite{Chaturantabut2016}, \cite{HesthavenReview2021}, \cite{LallMarsden2003}.

\subsection{Outline}
The main content of the remainder of this paper is, as follows. In Section~\ref{sec: Model reduction} we review several model reduction techniques and set the appropriate notation. In Section~\ref{sec: Numerical experiment} we present the results of our numerical experiment demonstrating the applicability of model reduction techniques to particle methods for the Vlasov equation. Section~\ref{sec:Summary} contains the summary of our work.

\section{Model reduction}
\label{sec: Model reduction}

In this section we briefly review several model reduction techniques and set the notation appropriate for the problem defined in the introduction.

\subsection{Proper orthogonal decomposition}
\label{sec: Proper orthogonal decomposition}

Proper orthogonal decomposition (POD) is one of the standard model reduction techniques (see \cite{Antoulas2001}, \cite{Moore1981}). Consider a general ODE

\begin{equation}
\label{eq: General ODE}
\dot u  = g(u), \qquad\qquad \text{with $g:\mathbb{R}^N \longrightarrow \mathbb{R}^N$},
\end{equation}

\noindent
and with the initial condition $u(0)=u_0$. Equation~\eqref{eq: ODEs for the particles} has this form with $N=2nd$, $u=(X,V)$, and $g(u) = (V_1,\ldots,V_n,\frac{\partial \phi}{\partial x}(X_1),\ldots,\frac{\partial \phi}{\partial x}(X_n))$. When $N$ is a very high number, as is typical for particle methods, the system \eqref{eq: General ODE} becomes very expensive to solve numerically. The main idea of model reduction is to approximate such a high-dimensional dynamical system using a lower-dimensional one that can capture the dominant dynamic properties. Let $\Delta$ be an $N\times M$ matrix representing empirical data on the system \eqref{eq: General ODE}. For instance, $\Delta$ can be a collection of snapshots of a solution of this system, 

\begin{equation}
\label{eq: Snapshots of the solution}
\Delta = [u(t_1) \,\, u(t_2) \,\, \ldots\,\, u(t_M)],
\end{equation}

\noindent
at times $t_1, \ldots, t_M$. These snapshots are calculated for some particular initial conditions or values of parameters that the system \eqref{eq: General ODE} depends on. A low-rank approximation of $\Delta$ can be done by performing a singular value decomposition (SVD) of $\Delta$ and truncating it after the first $K$ largest singular values, that is,

\begin{equation}
\label{eq: SVD of Delta}
\Delta = U \Sigma V^T \approx U_K \Sigma_K V_K^T,
\end{equation}

\noindent
where $\Sigma = \text{diag}(\sigma_1, \sigma_2, \ldots)$ is the diagonal matrix of the singular values, $U$ and $V$ are orthogonal matrices, $\Sigma_K$ is the diagonal matrix of the first $K$ largest singular values, and $U_K$ and $V_K$ are orthogonal matrices constructed by taking the first $K$ columns of $U$ and $V$, respectively. Let $\xi$ denote a vector in $\mathbb{R}^K$. Substituting $u = U_K \xi$ in \eqref{eq: General ODE} yields a reduced ODE for $\xi(t)$ as

\begin{equation}
\label{eq: Reduced ODE}
\dot \xi = U_K^T g(U_K \xi),
\end{equation}

\noindent
with the projected initial condition $\xi(0)=\xi_0=U_K^T u_0$. If the singular values of $\Delta$ decay sufficiently fast, then one can obtain a good approximation of $\Delta$ for $K$ such that $K\ll N$ (see Section~\ref{eq: Approximation of data by linear subspaces}). Equation \eqref{eq: Reduced ODE} is then a low-dimensional approximation of \eqref{eq: General ODE} and can be solved more efficiently. For more details about the POD method we refer the reader to \cite{Antoulas2001}. In the context of particle methods for the Vlasov equation, the reduced model \eqref{eq: Reduced ODE} allows one to perform numerical computations with a much smaller number of degrees of freedom. Note, however, that while \eqref{eq: ODEs for the particles} is a Hamiltonian system, there is no guarantee that the reduced model \eqref{eq: Reduced ODE} will also have that property. In Section~\ref{sec: Numerical experiment} we demonstrate that retaining the Hamiltonian structure in the reduced model greatly improves the quality of the numerical solution. 

\subsection{Proper symplectic decomposition}
\label{sec: Proper symplectic decomposition}

Note that the Hamiltonian system \eqref{eq: Hamiltonian ODEs for the particles} can be equivalently written as

\begin{equation}
\label{eq: Hamiltonian system in terms of J}
\dot u = \mathbb{J}_{2N} \nabla_u H(u),
\end{equation}

\noindent 
where $u=(X,V)$ and $N=nd$. A model reduction technique that retains the symplectic structure of Hamiltonian systems was introduced in \cite{PengMohseni2016}. In analogy to POD, this method is called the proper symplectic decomposition (PSD). A $2N \times 2K$ matrix is called symplectic if it satisfies the condition

\begin{equation}
\label{eq: Symplectic matrix}
A^T \mathbb{J}_{2N} A = \mathbb{J}_{2K}.
\end{equation}

\noindent
For a symplectic matrix $A$, we can define its symplectic inverse $A^+ = \mathbb{J}_{2K}^T A^T \mathbb{J}_{2N}$. It is an inverse in the sense that $A^+ A = \mathbb{I}_{2K}$. Let $\xi$ be a vector in $\mathbb{R}^{2K}$. Substituting $u = A\xi$ in \eqref{eq: Hamiltonian system in terms of J} yields a reduced equation

\begin{equation}
\label{eq: Reduced Hamiltonian system}
\dot \xi = A^+ \mathbb{J}_{2N} \nabla_u H(u) = \mathbb{J}_{2K} \nabla_\xi H(A \xi),
\end{equation}

\noindent
which is a lower-dimensional Hamiltonian system with the Hamiltonian $\tilde H(\xi) = H(A \xi)$. Given a set of empirical data on a Hamiltonian system, the PSD method constructs a symplectic matrix $A$ which best approximates that data in a lower-dimensional subspace. We have tested two PSD algorithms, namely the cotangent lift and complex SVD algorithms.

\subsubsection{Cotangent lift algorithm}
\label{sec: Cotangent lift algorithm}
This algorithm constructs a symplectic matrix $A$ which has the special block diagonal structure

\begin{equation}
\label{eq: Cotangent lift matrix A}
A = \left( \begin{matrix}
\Phi & 0 \\
0 & \Phi
\end{matrix} \right),
\end{equation}

\noindent
where $\Phi$ is an $N \times K$ matrix with orthogonal columns, i.e. $\Phi^T \Phi = \mathbb{I}_{K}$. Suppose snapshots of a solution are given as an $N \times 2M$ matrix $\Delta$ of the form

\begin{equation}
\label{eq: Snapshots of the solution for cotangent lift}
\Delta = [X(t_1)\,\,\ldots \,\,X(t_M)\,\,V(t_1)\,\,\ldots\,\,V(t_M)].
\end{equation}

\noindent
The SVD of $\Delta$ is truncated after the first $K$ largest singular values, similar to \eqref{eq: SVD of Delta}. The matrix $\Phi$ is then chosen as $\Phi=U_K$. More details can be found in \cite{PengMohseni2016}.

\subsubsection{Complex SVD algorithm}
\label{sec: Complex SVD algorithm}

By allowing a broader class of symplectic matrices we may get a better approximation. The complex SVD algorithm constructs a symplectic matrix of the form

\begin{equation}
\label{eq: Complex SVD matrix A}
A = \left( \begin{matrix}
\Phi & -\Psi \\
\Psi & \Phi
\end{matrix} \right),
\end{equation}

\noindent
where $\Phi$ and $\Psi$ are $N \times K$ matrices satisfying the conditions

\begin{equation}
\label{eq: Conditions for Phi and Psi}
\Phi^T \Phi + \Psi^T \Psi = \mathbb{I}_K, \qquad\qquad \Phi^T \Psi = \Psi^T \Phi.
\end{equation}

\noindent
Suppose snapshots of a solution are given as an $N \times M$ complex matrix $\Delta$ of the form

\begin{equation}
\label{eq: Snapshots of the solution for complex SVD}
\Delta = [X(t_1)+i V(t_1)\,\,\ldots \,\,X(t_M)+i V(t_M)], 
\end{equation}

\noindent
where $i$ denotes the imaginary unit. The complex SVD of $\Delta$ is truncated after the first $K$ largest singular values, that is,

\begin{equation}
\label{eq: SVD of Delta for complex SVD}
\Delta = U \Sigma V^\dagger \approx U_K \Sigma_K V_K^\dagger.
\end{equation}

\noindent
The matrices $\Phi$ and $\Psi$ are then chosen as the real and imaginary parts of $U_K$, respectively, that is, $U_K = \Phi + i \Psi$. More details can be found in \cite{PengMohseni2016}.

%

\subsubsection{Preservation of the Hamiltonian}
\label{sec: Preservation of the Hamiltonian}

Let $u(t)$ be a solution of \eqref{eq: Hamiltonian system in terms of J} with the initial condition $u_0$, and let $\xi(t)$ be a solution of the reduced system \eqref{eq: Reduced Hamiltonian system} with the projected initial condition $\xi_0 = A^+u_0$. Since the solutions of both the unreduced and reduced equations preserve their respective Hamiltonians, we have that the error of the Hamiltonian $\Delta H(t) = H(u(t)) - \tilde H(\xi(t))$ arising due to the symplectic projection is constant in time and equal to its initial value $\Delta H(0) = H(u_0) - \tilde H(\xi_0)$, whose magnitude can be controlled by choosing a sufficiently high $K$. If geometric integrators are used to solve the reduced model \eqref{eq: Reduced Hamiltonian system}, then the numerical values of the reduced Hamiltonian $\tilde H$ stay very close to the exact value $\tilde H(\xi_0)$, and therefore they also stay close to the exact energy $H(u_0)$ of the unreduced system \eqref{eq: Hamiltonian system in terms of J}.

\subsection{Approximation of data by linear subspaces}
\label{eq: Approximation of data by linear subspaces}

The POD and PSD methods described in Sections~\ref{sec: Proper orthogonal decomposition} and~\ref{sec: Proper symplectic decomposition}, respectively, rely on the assumption that the set of the empirical data $\Omega=\{ u(t_1),\ldots,u(t_M)\}\subset \mathbb{R}^N$ can be approximated well by a linear subspace of dimension $K<N$, otherwise these techniques do not bring computational savings. The so called Kolmogorov $n$-width $d_K(\Omega)$ describes the error arising from the projection of $\Omega$ onto the best-possible subspace of $\mathbb{R}^N$ of a given dimension $K$ (see \cite{PinkusBook1985}). It has been proven that for a class of problems written as parametrized PDEs the Kolmogorov $n$-width decays exponentially fast, that is, $d_K(\Omega) = O(e^{-\gamma K})$ with some constant $\gamma>0$ (see \cite{BuffaMaday2012}, \cite{Ohlberger2016}). This extremely fast decay plays a critical role in any model reduction strategy based upon projecting to linear subspaces, since it allows one to select a low to moderate $K$ to achieve small approximation errors. Such theoretical results have not yet been proven for particle-based simulations of the Vlasov equation. We will show by performing explicit numerical computations that for the example presented in Section~\ref{sec: Numerical experiment} the evolution of the particles is indeed well approximated in a low dimensional subspace. In case the empirical data do not appear to lie in a linear subspace, instead of the POD and PSD methods described above, one may apply online adaptive methods that update local reduced spaces depending on time (\cite{CarlbergAdaptive2015}, \cite{Peherstorfer2015}), as well as a structure-preserving dynamic reduced basis method for Hamiltonian systems (\cite{Pagliantini2021}). The application of the latter two techniques to the Vlasov equation will be investigated in a follow-up work.

\section{Numerical experiment}
\label{sec: Numerical experiment}

In this section we present the results of a simple numerical experiment demonstrating the applicability of model reduction techniques to particle methods for the Vlasov equation.

\subsection{Initial and boundary conditions}
\label{eq: Initial and boundary conditions}

We consider the Vlasov equation \eqref{eq: Vlasov equation} on a one-dimensional ($d=1$) spatial domain $-\infty\leq x \leq \infty$ with the initial condition

\begin{equation}
\label{eq: Initial condition for the particle density function}
f(0,x,v) = f_0(x,v)=  \frac{1}{\sqrt{2 \pi} \eta} e^{-\frac{1}{2\eta^2}x^2} \bigg( \frac{1}{1+a}\frac{1}{\sqrt{2 \pi}} e^{-\frac{1}{2}v^2} + \frac{a}{1+a}\frac{1}{\sqrt{2 \pi} \sigma} e^{-\frac{1}{2\sigma^2}(v-v_0)^2}   \bigg),
\end{equation}

\noindent
where the parameters are set as follows:

\begin{equation}
\label{eq: Parameter values}
\eta=10, \qquad a=0.3, \qquad v_0=4, \qquad \sigma=0.5.
\end{equation}

\noindent
This is a ``bump-on-tail'' distribution in velocity space combined with a Gaussian distribution in position space. The initial conditions for the particle positions $X_i(0)$ and velocities $V_i(0)$ in \eqref{eq: ODEs for the particles} are generated as random variables drawn from the probability distribution \eqref{eq: Initial condition for the particle density function} using rejection sampling.

\subsection{Empirical data}
\label{eq: Empirical data}

Let the Vlasov equation \eqref{eq: Vlasov equation} be parameter-dependent. For example, the external electric field may depend on some parameter $\beta$, i.e., $E=E(x;\beta)$. Suppose we have the following computational problem: we would like to scan the domain of $\beta$, that is, compute the numerical solution of \eqref{eq: Vlasov equation} for a large number of values of $\beta$. Given that in practical applications the system \eqref{eq: ODEs for the particles} is very high-dimensional, this task is computationally very intensive. Model reduction can alleviate this substantial computational cost. One can carry out full-scale computations using high-fidelity numerical methods only for a selected small number of values of $\beta$. These data can then be used to identify reduced models, as described in Section~\ref{sec: Model reduction}. The lower-dimensional equations \eqref{eq: Reduced ODE} or \eqref{eq: Reduced Hamiltonian system} can then be solved more efficiently for other values of $\beta$, thus reducing the overall computational cost. For our simple experiment, we consider a linear external electric field, namely

\begin{equation}
\label{eq: Electric field}
E(x;\beta) = \beta^2 x,
\end{equation}

\noindent
where $\beta$ is a real parameter. While this is a rather academic example, it allows an easy demonstration of how model reduction works in the case of particle methods for the Vlasov equation. With this electric field one can solve the Vlasov equation \eqref{eq: Vlasov equation} using the method of characteristics to obtain the exact solution

\begin{equation}
\label{eq: Exact solution of the Vlasov equation}
f(t,x,v) = f_0\bigg(x \cos \beta t - \frac{1}{\beta}v \sin \beta t, \beta x \sin \beta t + v \cos \beta t\bigg)
\end{equation}

\noindent
satisfying the initial condition \eqref{eq: Initial condition for the particle density function}. Moreover, the equation for the trajectories of the particles \eqref{eq: ODEs for the particles} is solved by

\begin{align}
\label{eq: Exact solution of the particle equations}
X_i(t) = \frac{1}{\beta}V_i(0) \sin \beta t + X_i(0) \cos \beta t, \qquad\qquad V_i(t) = V_i(0) \cos \beta t - \beta X_i(0) \sin \beta t.
\end{align}

\noindent
Instead of solving the full-scale system numerically, for convenience we used the exact solution \eqref{eq: Exact solution of the particle equations} to generate the empirical data for the following six values of the parameter $\beta$:

\begin{equation}
\label{eq: beta values}
\beta_1 = 5.95, \quad \beta_2 = 5.97, \quad \beta_3 = 5.99, \quad \beta_4 = 6.01, \quad \beta_5 = 6.03, \quad \beta_6 = 6.05.
\end{equation}

\noindent
The exact solution \eqref{eq: Exact solution of the particle equations} was sampled for $n=1000$ particles at times $t_k = k\Delta t$ for $\Delta t = 0.01$ and $k=0, 1, \ldots, 100000$, i.e., over the time interval $0\leq t \leq 1000$. It should be noted that the same initial values for the positions and velocities of the particles were used for each value of $\beta$. The generated data for all values of $\beta$ were put together and used to form the snapshot matrices \eqref{eq: Snapshots of the solution}, \eqref{eq: Snapshots of the solution for cotangent lift}, and \eqref{eq: Snapshots of the solution for complex SVD}. For instance, for the POD snapshot matrix \eqref{eq: Snapshots of the solution} we used

\begin{equation}
\label{eq: Snapshots of the solution in the experiment}
\Delta = [u(t_0;\beta_1) \,\, u(t_1;\beta_1) \,\, u(t_2;\beta_1) \,\, \ldots \,\, u(t_0;\beta_2) \,\, u(t_1;\beta_2)\,\, u(t_2;\beta_2) \,\, \ldots].
\end{equation}

\begin{figure}[tbp]
	\centering
		\includegraphics[width=.8\textwidth]{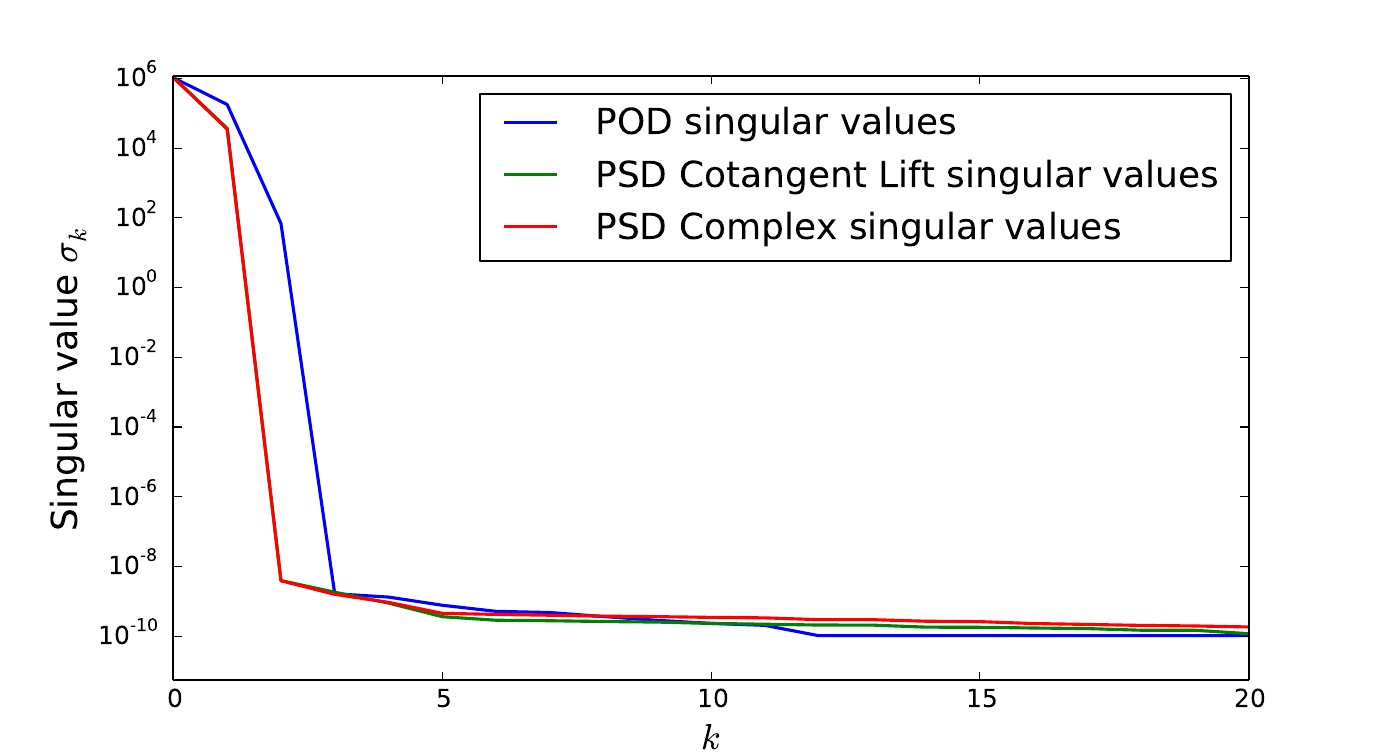}
		\caption{ The decay of the singular values as calculated for the empirical data with each of the SVD-based algorithms. }
		\label{fig: Singular values}
\end{figure}

\noindent
Then, following the description of each algorithm in Section~\ref{sec: Model reduction}, reduced models were derived. The decay of the singular values for the POD, PSD cotangent lift and complex SVD methods is depicted in Figure~\ref{fig: Singular values}. Since the sizes of the snapshot matrices $\Delta$ in our experiment were not exceedingly high (e.g., $2000\times 600006$ for \eqref{eq: Snapshots of the solution in the experiment}), we computed the full SVD decompositions using the standard SVD algorithm implemented in the Julia programming language. This algorithm requires memory and time that are superlinear in the dimensions of $\Delta$ (see \cite{Drineas2006}), which is prohibitive for very large data sets. However, the purpose of model reduction is to determine only a small number of the largest singular values and their corresponding singular vectors, therefore algorithms such as the truncated (see \cite{ChanTSVD1990}) or the randomized (see \cite{HaikoTroppSVD2011}) SVD decompositions can be used instead. These algorithms require significantly less memory and time than the full SVD decomposition. Overall, this \emph{offline} stage of model reduction is expensive, but it is performed only once. Then, in the \emph{online} stage, the reduced systems can be solved cheaply for an arbitrary number of values of $\beta$.

\subsection{Reduced model simulations}
\label{eq: Reduced model simulations}

To test the accuracy of the considered model reduction methods, we have compared the results of reduced model simulations to a full-scale reference solution. The reference solution for $\beta=6.0$ was calculated on the time interval $0 \leq t \leq 1000$ in the same way as the empirical data in Section~\ref{eq: Empirical data}. Note that for this choice of $\beta$ the period of the reference solution is $T\approx 1.05$ (see \eqref{eq: Exact solution of the particle equations}), and the considered time interval encompasses roughly 955 periods. The reduced models were solved numerically on the same time interval using the second-order explicit and implicit midpoint methods. Note that when applied to a Hamiltonian system, the implicit midpoint method is a symplectic integrator, while the explicit midpoint method is not (see, e.g., \cite{HLWGeometric}). All simulations were carried out with the time step $\Delta t = 0.0001$. The POD model \eqref{eq: Reduced ODE} was solved for $K=5$ and $K=10$ (thus reducing the dimensionality of the problem from $2n=2000$ to 5 and 10, respectively). Similarly, the PSD model \eqref{eq: Reduced Hamiltonian system} was solved for $K=5$ and $K=10$, both for the cotangent lift and complex SVD algorithms, in both cases reducing the dimensionality to 10 and 20, respectively. The choice of $K$ is a compromise between the speed and the accuracy: the smaller $K$ the faster the computation, but also the larger the projection error. In practice, one may choose $K$ based on the initial value of the error \eqref{eq: Relative error}, i.e., the value of the projection error for the initial condition. For instance, in our experiment, the initial relative error for the POD simulations was equal to $1.09\cdot 10^{-14}$ for $K=5$, and $7.81\cdot 10^{-15}$ for $K=10$. All computations were performed in the Julia programming language with the help of the \emph{GeometricIntegrators.jl} library (see \cite{KrausGeometricIntegrators}). The three main conclusions from the numerical experiments are described below.

\subsubsection*{Long-time instability of the POD simulations}

\begin{figure}[tbp]
	\centering
		\includegraphics[width=.8\textwidth]{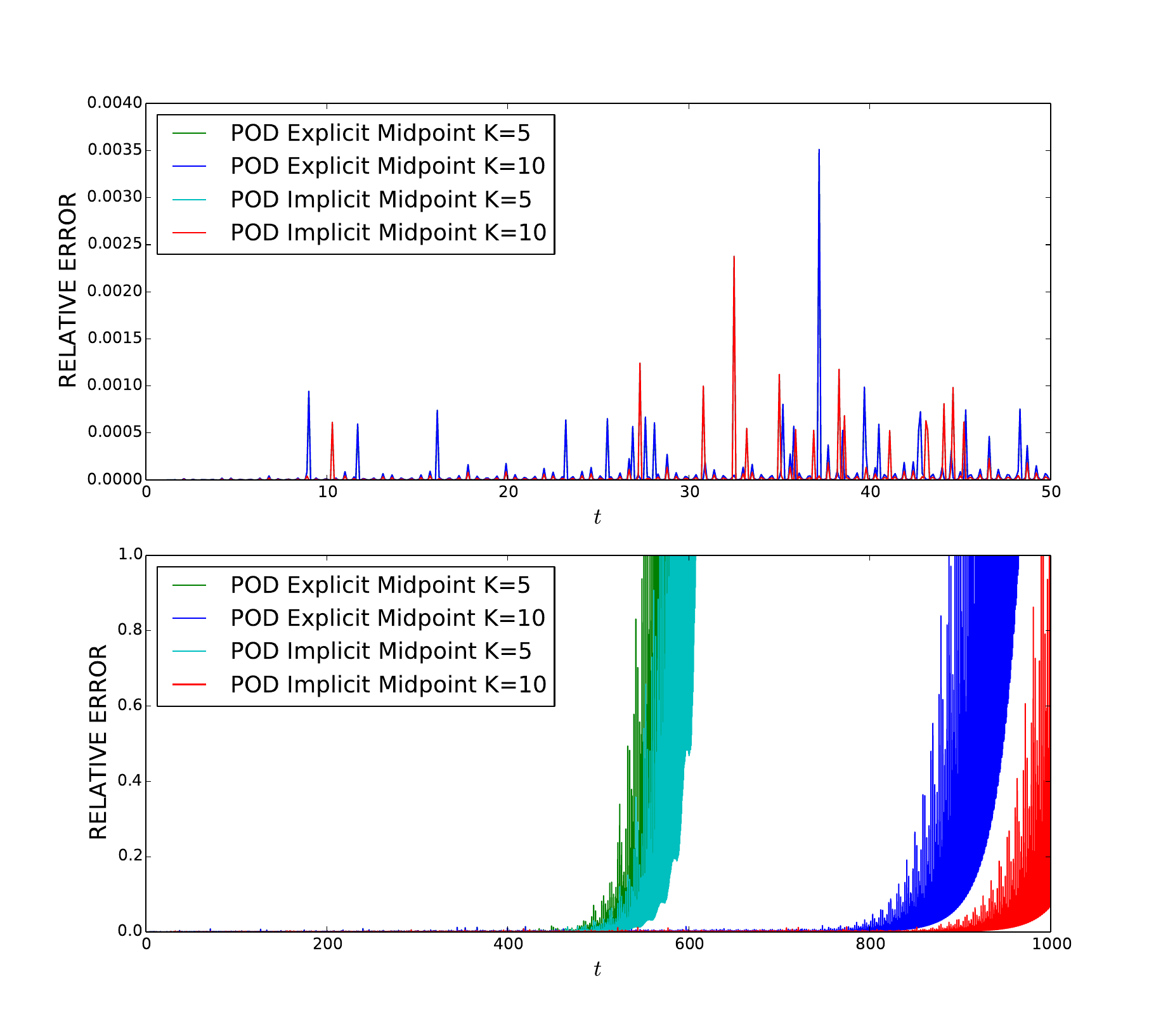}
		\caption{ \emph{Top:} The relative error of the POD simulations on the time interval $0\leq t \leq 50$. Both the explicit and implicit midpoint methods yield accurate approximations. Note that the plots for the explicit midpoint method with $K=5$ and $K=10$, as well as for the implicit midpoint method with $K=5$ and $K=10$, overlap very closely and are therefore indistinguishable. \emph{Bottom:} The same plot over the whole simulation interval $0\leq t \leq 1000$. It is evident that over a long integration time the errors blow up and both the explicit and implicit method simulations become unstable.}
		\label{fig: Error for POD}
\end{figure}

As a measure of accuracy of the reduced models we take the relative error

\begin{equation}
\label{eq: Relative error}
\frac{\| u(t)-u_{\text{ref}}(t)\|}{\| u_{\text{ref}}(t)\|},
\end{equation}

\noindent
where $u_{\text{ref}}$ is the reference solution of \eqref{eq: General ODE}, as described above, and $u = U_K \xi$ is the reconstructed solution, with $\xi$ being the numerical solution of the reduced model \eqref{eq: Reduced ODE}. The relative error as a function of time is depicted in Figure~\ref{fig: Error for POD}. We see that the POD simulations give very accurate results on shorter time intervals, but the errors blow up over a long integration time, and both the explicit and implicit midpoint method simulations become unstable. This is a consequence of the fact that there is no guarantee that the reduced system \eqref{eq: Reduced ODE} retains any stability properties of the original system \eqref{eq: General ODE}. In fact, the reduced equation for our example takes the form of the linear equation $\dot \xi = \Lambda_K \xi$, where the $K\times K$ matrix $\Lambda_K$ is given by

\begin{equation}
\label{eq: Lambda matrix}
\Lambda_K = U_K^T \left(\begin{matrix}
0 & \mathbb{I}_{n} \\
-\beta^2 \mathbb{I}_{n} & 0
\end{matrix}\right)  U_K.
\end{equation}

\noindent
In our experiment, for $K=5$ the matrix $\Lambda_K$ has 5 eigenvalues with positive real parts, the largest one of which equals $\text{Re}\, \lambda \approx 0.0655$. Similarly, for $K=10$ the eigenvalue with the largest real part is $\text{Re}\,\lambda \approx 0.0403$. The modes corresponding to these eigenvalues grow exponentially, and after a certain amount of time dominate the solution. This means that the reduced system is unstable; therefore, the errors arising from projecting the initial condition $\xi_0 = U_K^T u_0$ and from applying numerical integration schemes amplify over the simulation time, eventually leading to the observed loss of accuracy (see \cite{Prajna2003}, \cite{RathinamPetzold2003}).

\subsubsection*{Long-time stability of the PSD simulations}

\begin{figure}[tbp]
	\centering
		\includegraphics[width=.8\textwidth]{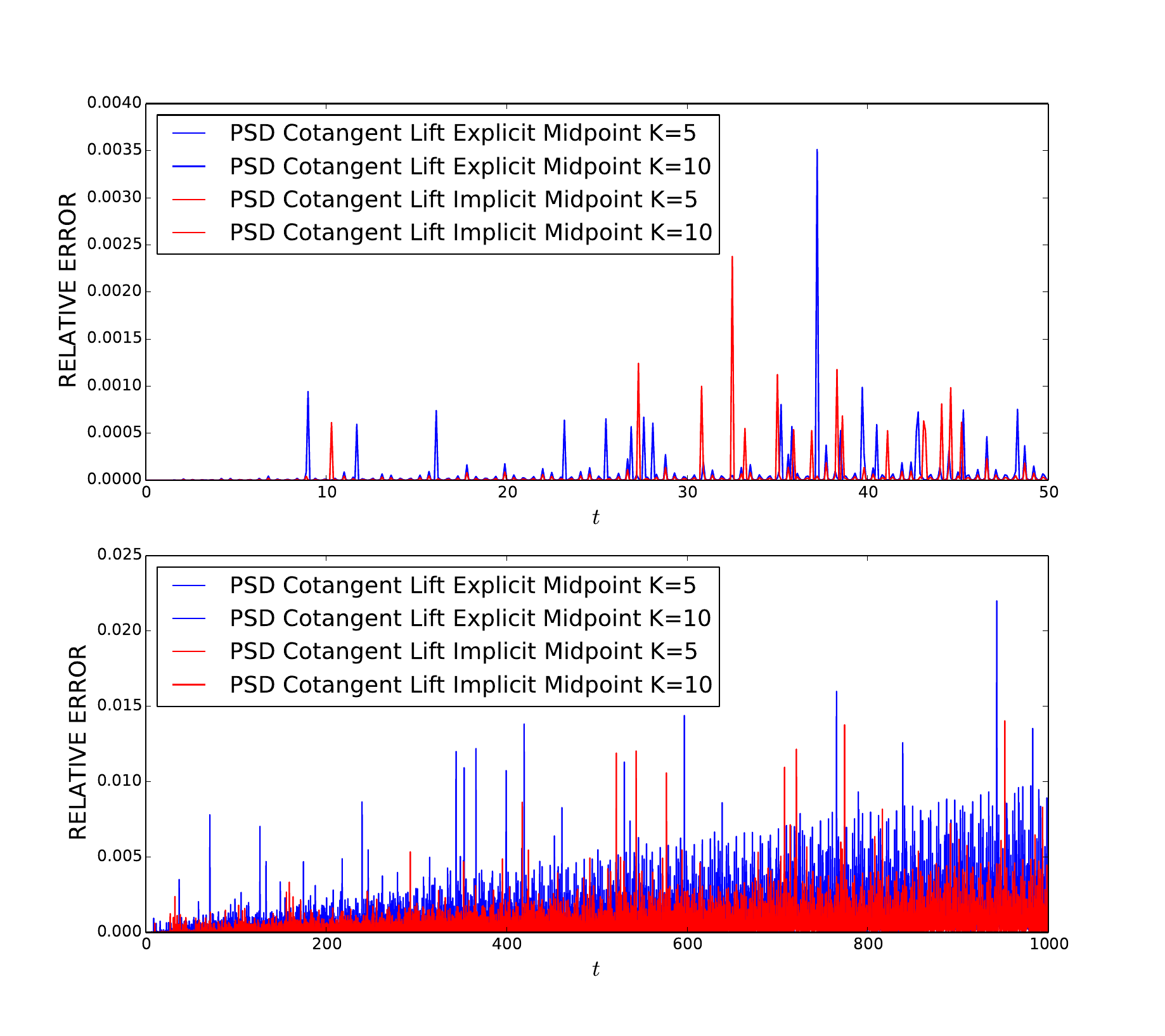}
		\caption{ The relative error of the PSD simulations carried out with the cotangent lift algorithm on the time intervals $0\leq t \leq 50$ (\emph{Top}) and $0\leq t \leq 1000$ (\emph{Bottom}). The errors on the interval $0\leq t \leq 50$ are essentially identical for the POD and PSD simulations, but the PSD simulations retain good accuracy and stability over the whole integration time. The results for the complex SVD algorithm are nearly identical. Note that the same color code is used for the plots that overlap very closely and are therefore indistinguishable.}
		\label{fig: Error for PSD}
\end{figure}

In the case of the PSD models the reconstructed solution $u=A \xi$ is obtained from the numerical solution $\xi$ of the reduced Hamiltonian system \eqref{eq: Reduced Hamiltonian system}. The relative error as a function of time for the cotangent lift algorithm is depicted in Figure~\ref{fig: Error for PSD}. We see that both the explicit and implicit midpoint methods retain good accuracy and stability over the whole integration time. It is therefore evident that by preserving the Hamiltonian structure of the particle equations, symplectic model reduction significantly improves the stability of the numerical computations even if a non-symplectic integrator (here the explicit midpoint method) is used. The numerical results for the complex SVD algorithm are nearly identical, therefore for brevity and clarity we skip presenting a separate figure.

\subsubsection*{Long-time energy behavior}

\begin{figure}[tbp]
	\centering
		\includegraphics[width=.8\textwidth]{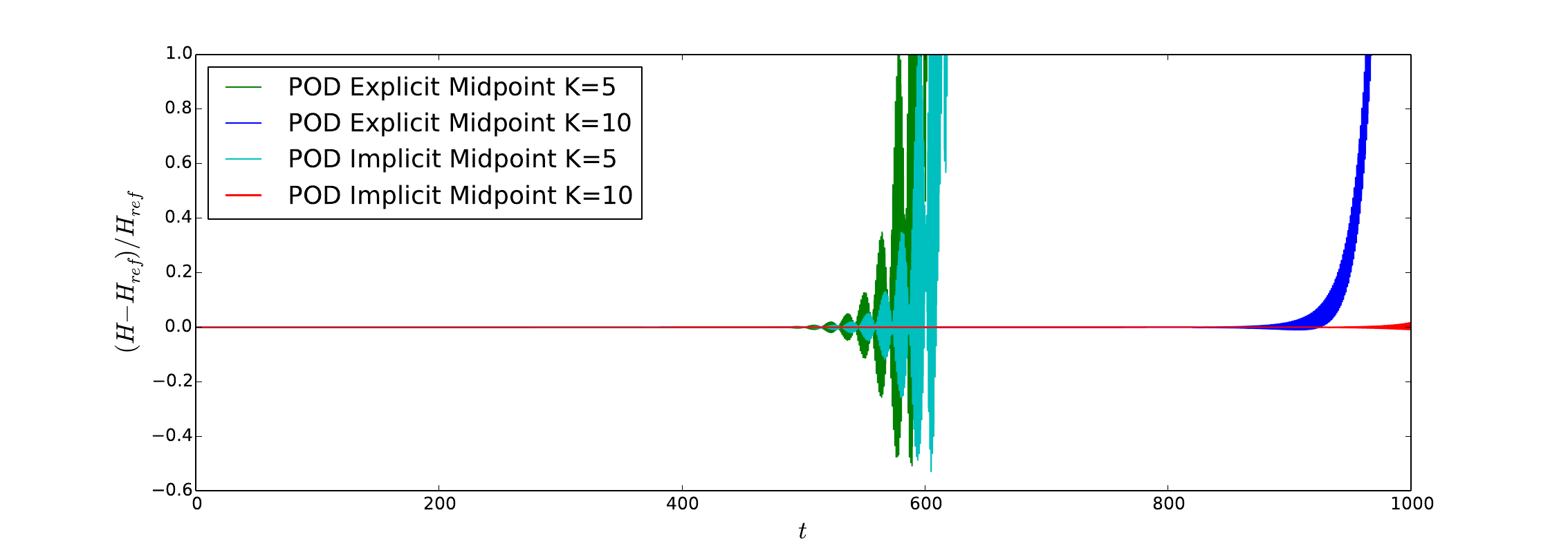}
		\caption{The relative error of the total energy of the particles as a function of time is depicted for the POD simulations. The energy eventually blows up for both the explicit and implicit midpoint methods.}
		\label{fig: EnergyPOD}
\end{figure}

\begin{figure}[tbp]
	\centering
		\includegraphics[width=.8\textwidth]{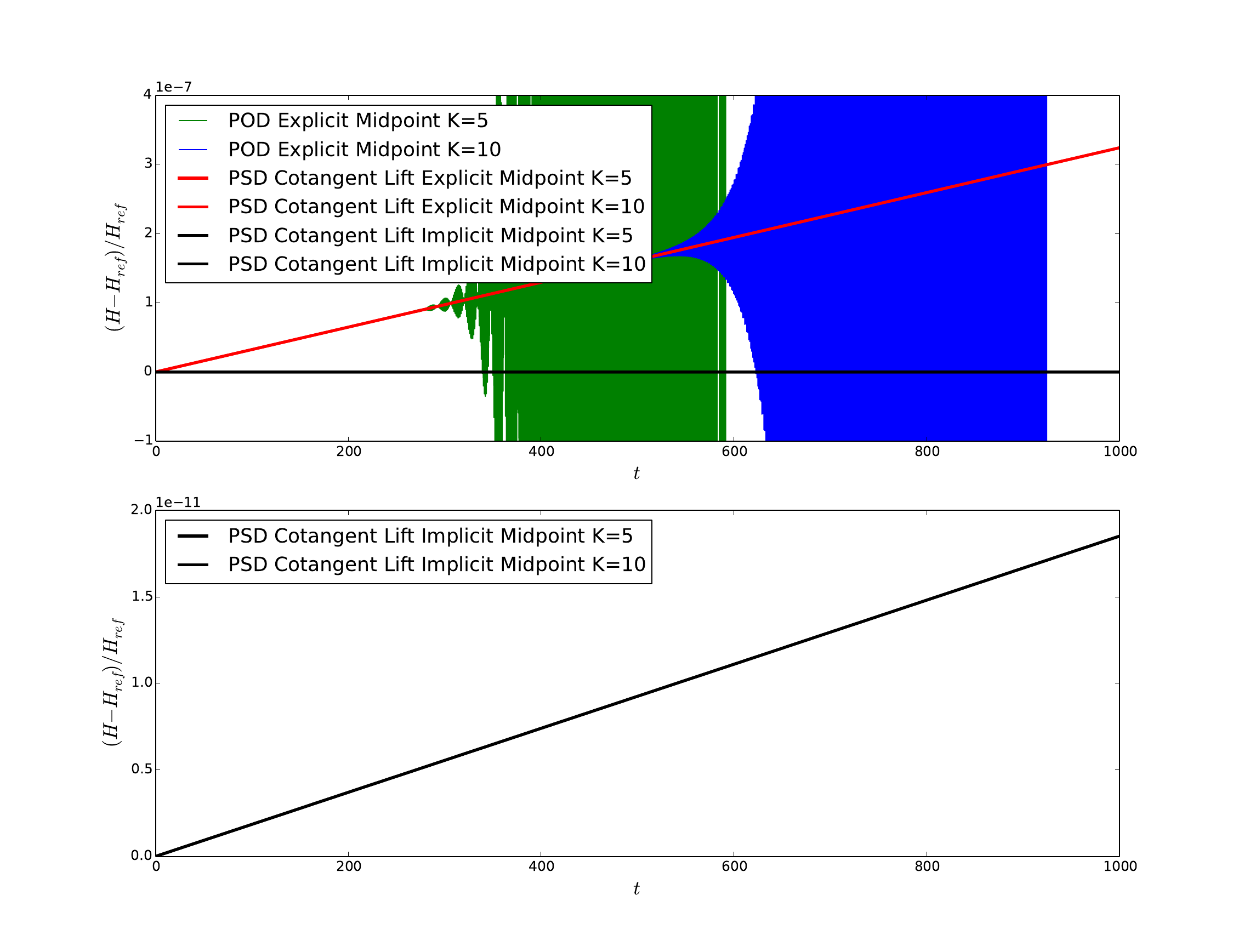}
		\caption{\emph{Top:} The relative error of the total energy of the particles as a function of time is depicted for the PSD simulations. The energy behavior is more stable than for the POD simulations. While the non-symplectic explicit midpoint method shows a linear growth trend, the symplectic implicit midpoint method appears to nearly preserve the total energy. For comparison, also two POD simulations are depicted. \emph{Bottom:} Same plot, but only the simulations for the symplectic implicit midpoint method are depicted in order to show the scale of energy preservation. In fact these simulations also show a linear trend, which is nevertheless four orders of magnitude smaller than for the non-symplectic explicit midpoint method. Note that the same color code is used for the plots that overlap very closely and are therefore indistinguishable. The results for the complex SVD algorithm are nearly identical, and are therefore omitted for clarity.}
		\label{fig: EnergyPSD}
\end{figure}

As the particle equations \eqref{eq: Hamiltonian ODEs for the particles} are Hamiltonian, the total energy \eqref{eq: Hamiltonian for the particles} of the particles should be preserved. In our numerical experiment the Hamiltonian for the reference solution \eqref{eq: Exact solution of the particle equations} with $\beta=6.0$ was $H_\text{ref}\approx 1.803\cdot 10^6$. The relative error of the total energy of the particles for each of the algorithms is depicted in Figure~\ref{fig: EnergyPOD} and Figure~\ref{fig: EnergyPSD}. One can clearly see that while the POD simulations initially retain the total energy relatively well, there is an evident linear growth trend for both the explicit and implicit midpoint methods, and the energy eventually blows up over a long integration time. On the other hand, the energy behavior for the PSD simulations is more stable. The non-symplectic explicit midpoint method applied to the cotangent lift algorithm also shows the same linear growth trend, but does not blow up. Furthermore, the implicit midpoint method, which is symplectic in this case, demonstrates near-preservation of the total energy, with only a minor linear growth throughout the whole simulation time. This demonstrates another important advantage of symplectic model reduction: since the reduced equations \eqref{eq: Reduced Hamiltonian system} are also Hamiltonian, one can employ symplectic time integrators to obtain a numerical solution that nearly conserves energy.

\section{Summary and future work}
\label{sec:Summary}

We have compared several model reduction techniques and demonstrated their usefulness for particle-based simulations of the Vlasov equation. We have pointed out the importance of retaining the Hamiltonian structure of the equations governing the evolution of particles. Our work can be extended in several directions.
First, model reduction methods can be applied to the Vlasov equation coupled to a self-consistent electric field satisfying the Poisson equation, or electromagnetic fields satisfying the Maxwell equations (see \cite{KrausGEMPIC}).
Second, model reduction of collisional Vlasov equations stemming from metriplectic brackets (see \cite{HirvijokiKrausMetriplectic}) or stochastic action principles (see \cite{KrausTyranowski2019}, \cite{TyranowskiVlasovMaxwell}, \cite{TyranowskiStochasticModelReduction}) would be an interesting and useful extension of the work presented here.

\section*{Acknowledgments}
We would like to thank Christopher Albert, Tobias Blickhan, Eric Sonnendr\"{u}cker and Udo von Toussaint for useful comments and references. The study is a contribution to the Reduced Complexity Models grant number ZT-I-0010 funded by the Helmholtz Association of German Research Centers.



\end{document}